# Portable neutron/gamma scintillation detector for status monitoring of accelerator-driven neutron source IREN

S. Nuruyev[a,b], D. Berikov[a,c*], R. Akbarov[a,b,d], G. Ahmadov[a,b,d,e], F. Ahmadov[b,d], A. Sadigov[b,d], M. Holik[f,j], J. Naghiyev[d], A. Madadzada[a,d], K. Udovichenko[a]

[a]Joint Institute for Nuclear Researches, Joliot-Curie 6, 141980 Dubna, Russia
[b]Institute of Radiation Problems under Ministry of Science and Education, B.Vahabzade str. 9, AZ1143 Baku, Azerbaijan
[c]Institute of Nuclear Physics of the National Nuclear Center of Kazakhstan, Ibragimova 1, 050032 Almaty, Kazakhstan
[d]Innovation and Digital Development Agency Nuclear Research Department, Gobu str. 20th km of Baku-Shamakhi Highway, AZ0100 Baku, Azerbaijan
[e]Innovative Electronics and Detectors LLC, Badamdard STQ-1, AZ1021 Baku, Azerbaijan
[f]Faculty of Electrical Engineering, UWB in Pilsen, Univerzitni 2795/26, 306 14 Pilsen, Czech Republic
[j]Institute of Experimental and Applied Physics, CTU in Prague, Husova 240/5, 110 00 Prague, Czech Republic

*Corresponding author.

E-mail address: daniyar.berikov@gmail.com

**ABSTRACT**

Accelerator-driven system (ADS) facilities world-wide opens new opportunities for nuclear physics investigations, so that a high flux of neutrons through spallation reactions can be produced at these facilities. It is known that the measurement, continuous monitoring and optimization of the particle accelerator beam intensity are among the most important actions in the operation of such facilities. Considering this point of view, this paper presents a neutron/gamma counter based on a micropixel avalanche photodiode (MAPD) and a plastic scintillator that monitors the status of the accelerator-driven intense resonance neutron source (IREN) facility by measuring the neutron-gamma intensity in the target hall. The electronics of the modular neutron counter has been designed and developed, including a bias voltage source (up to 130 V), a preamplifier (36 gain) and discriminator (>10 mV) circuitry. The last product of MAPD (operation voltage- 55 V, PDE- 33 %, total number of pixels- 136900) was used as a photon readout from a plastic scintillator. The sensitive area of MAPD was $3.7 \times 3.7$ mm$^2$ and the size of the plastic scintillator $3.7 \times 3.7 \times 30$ mm$^3$. The measurement was carried out in the IREN target hall, where it was necessary to monitor not only high neutron fluxes, but also gamma quanta. The experimental results demonstrated a dependence between the count rate of the detector and the frequency of the accelerator, which ranges from 2 to 50 Hz.

## 1. Introduction

Accelerator-based neutron sources have become a popular alternative to research reactors for neutron beam research over the past few decades, which provide the most intense pulsed neutron beams. These facilities use accelerated particles that hit a heavy metal target, such as lead or tungsten, which produces a high flux of neutrons. One of these systems is the IREN facility. The IREN facility is an intense pulsed source of resonant neutrons designed for nuclear physics research using the time-of-flight method in the neutron energy range up to hundreds of keV and for studying photonuclear reactions. This is a combination of a linear electron accelerator (LUE-200 with a beam power about 10 kW,) and a subcritical multiplying target (tungsten based alloy). Electrons from the accelerator induce bremsstrahlung quanta in the target, which, in turn, induce photonuclear reactions (γ, xn) [1]. A more detailed description, operation principle, main characteristics and parameters of the IREN neutron source are given in [1-3].

As known, when conducting nuclear physics research using neutron sources, the accuracy of the obtained results is affected by two parameters of the facility – the stability of the neutron flux over time and the effect/background ratio. These parameters depend on the operating mode of the facility. To adjust the optimal parameters of the neutron source and maintain them in a stable state, a device is needed that allows these parameters to be displayed in real time. Such devices include neutron/gamma monitors. There are several basic approaches to monitor neutrons including scintillation detectors, gas detectors, solid-state detectors etc. Each of these approaches has its advantages and limitations, and the choice of detector depends on the specific application and the required sensitivity, resolution, count rate performance and range of detection energy. It has been a challenge for neutron sensor providers over the past decade to find suitable replacements for the widely used $^3$He detectors. Sensors based on $^3$He, which are offered by manufacturers worldwide, represent the state of the art in neutron detection. But, $^3$He world crisis requires a development of new methods of neutron detection to replace commonly used $^3$He proportional counters [4].

Radiation detectors based on silicon photomultiplier (SiPM) coupled to a scintillator are becoming increasingly popular, gradually replacing traditional photomultiplier tube (PMT) based detectors and considered as promising candidate to replace $^3$He detectors [4-6]. They are widely used in many fields of science and technology such as high energy physics, medical imaging, space research, dosimetry, and security [6–12]. Its compactness, relatively low bias voltage, cost, and immunity to electromagnetic interference (EMI) give the SiPM superiority over traditional PMTs in radiation detector implementations [13-16]. Some of these detectors have already been implemented in serial or pilot production, many are known only from individual publications.

This paper presents a detector module based on a silicon photomultiplier and an organic scintillator, which can be used instead of traditional neutron/gamma monitors, which makes it possible to increase their sensitivity and simplify the recording equipment. The developed detector module makes it possible to control the relative neutron flux in real time. The detector module was successfully tested in the target hall of the resonant neutron source IREN (JINR, Dubna).

## 2. Detector module

Scintillation detectors typically consist of a scintillation crystal, a photodetector, and an electronic system. The scintillation crystal is the part that interacts with incident particles, such as neutrons or gamma rays, and produces light as a result of the interaction. The use of plastic scintillation detectors is considered advantageous for neutron detection due to their convenience and practicality. The unlimited volume and configuration of such a detector provide high efficiency of neutron detection and fast response time. Therefore, we have chosen inexpensive plastic scintillators so that we needed only counting neutrons/gammas. A plastic scintillator with 10000 photons/1 MeV light yield was purchased from Epic Crystal [17]. Its scintillation light has a short decay time (15-40 ns) and peaks in the blue region with wavelength $390 \div 430$ nm, which matches the highest spectral sensitivity of MAPD . The size of the plastic scintillator is $3.7 \times 3.7 \times 30$ mm$^3$.

Scintillation lights were registered by MAPD-3NM-2LOT photomultiplier manufactured by Zecotek Photonics Inc. In 2021 [18-19]. MAPD-3NM-2LOT has deeply buried pixel structure with the pixel size of 12 μm and the active area of $3.7 \times 3.7$ mm$^2$. Photon detection efficiency (PDE) of the photomultiplier is 30-35 %, avalanche gain is $2 \cdot 10^5$ at 3V overvoltage (or at the optimal voltage), the breakdown voltage is 52 V at room temperature (25 $^0$C). One can get familiar with a more detailed description of all generations of MAPDs in [20-22]. A top view of the electronic read-out and plastic scintillator with combination MAPD is shown in fig. 1. The plastic scintillator was covered with white teflon tape, except for one side, which is coupled to MAPD with a special liquid (a silicone optical grease) to reduce photon loss.

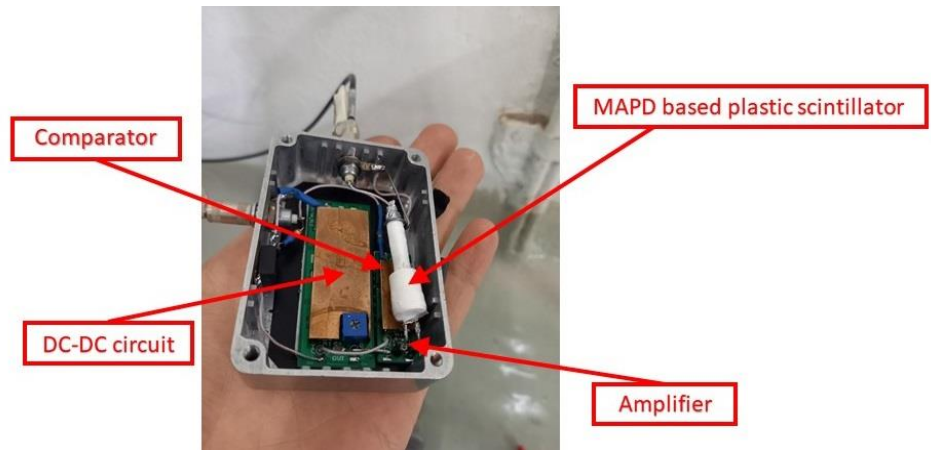

**Fig. 1.** Photo of the electronic system and the detector based on MAPD and plastic scintillator.

*2.1. Electronic parts of the detector module*

    The detector module includes the following electronic parts: DC-DC converter, preamplifier and comparator, which, in turn, is fed from the same positive 6 V source. The main part of the electronic circuit is the DC-DC converter for biasing MAPD. The MC34063A voltage controller is used to assemble the DC-DC converter. The converter allows converting the +3÷6 V voltage up to 90 V. The average fluctuation of the voltage output value is less than 0.01 V. One of the MC34063A channels is used as a pulse generator. Low-capacity capacitors (30 pF) are used to determine the frequency of the generated signal. The frequency of the signal was selected 10500 Hs, so that at this frequency, the current demand of the controller was minimum. The pulse has a rectangular shape, the frequency is 15000 Hs (resonance frequency-LC), the amplitude is 3V, and the duration is 30 ms. These pulses are fed to the coil with an inductance of 3700 µH. The energy collected in the inductive coil is discharged to the capacitor with a capacity of 10 nF via the diode BAV.102 (operating voltage 500 V). The discharge of the charge accumulated in the capacitor causes the output voltage to increase again and again. A semiconductor diode was employed to prevent reversal charge. Fig. 2 shows the high voltage DC-DC circuit diagram. The current demand of this type of voltage converter taken from the input was 2 mA. This type of DC-DC converter is intended to be used to bias MAPD photodetectors.

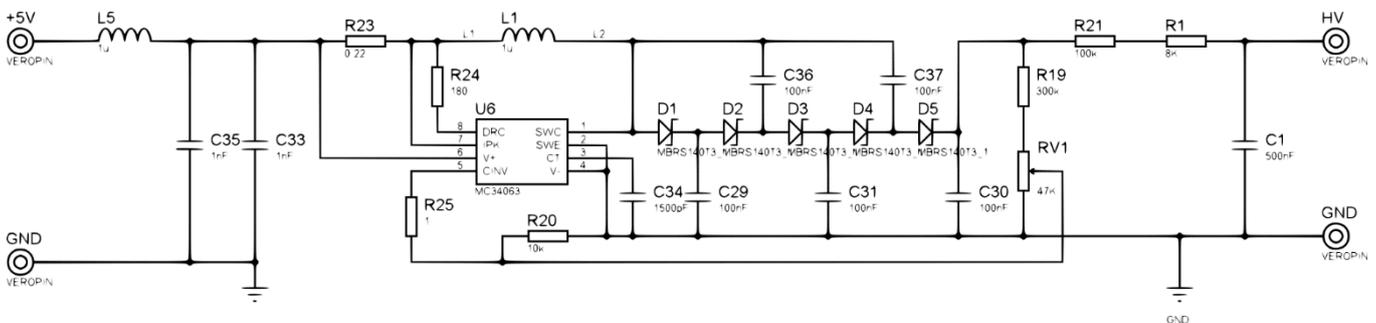

**Fig. 2.** DC-DC circuit diagram.

    In addition to the DC-DC converter, a preamplifier with 36 gain and a bandwidth of about 250 MHs was designed to amplify output analog signal from MAPD, since the duration of scintillation photons emitted from plastic scintillators varies in the range of 5 ns to 40 ns. The output signal of the preamplifier was an analog negative signal. The preamplifier was designed based on two transistors of type BFS17 and BFT92. The last part of the circuit is a signal comparator assembled using an LT-1355 operational amplifier. The threshold level for input signal was selected using a potentiometer.

    Variations in the output signal amplitude of the comparator were noticed as the signal amplitude changed. The signal received at the output of the comparator was fed to the base of the transistor. After the transistor, the output signals were observed at a constant amplitude. In addition, a 74HC74 flip-flop was added to the circuit to prevent the signal width changing. After the flip-flop, the signal received at the digital output had a constant width and a constant amplitude. Thus, there are two signal outputs: analog and TTL output from the electrical circuit.

## 3. The operation principle of the measuring equipment
The block diagram of the experimental setup is shown in fig. 3. MAPD with a scintillator, electronic components were placed inside the light-tight duralumin box with size of 11×6×3 cm (see fig. 1).

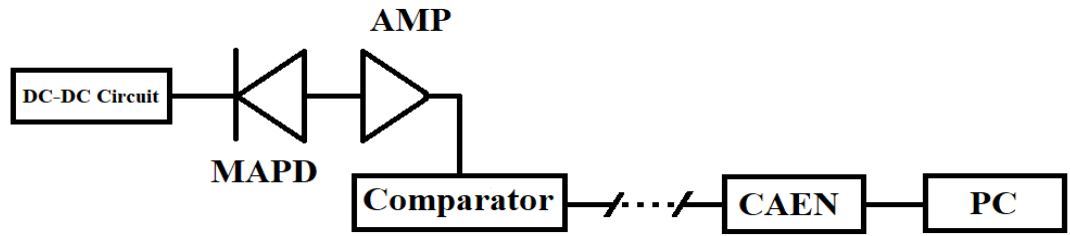

**Fig. 3.** Block diagram of the experimental setup.

The box was placed in the IREN target hall at a distance of 4 m from the center of the neutron generating target. It was on the floor (in height – 2 meters to the target) behind the protection of lead bricks (fig. 4).

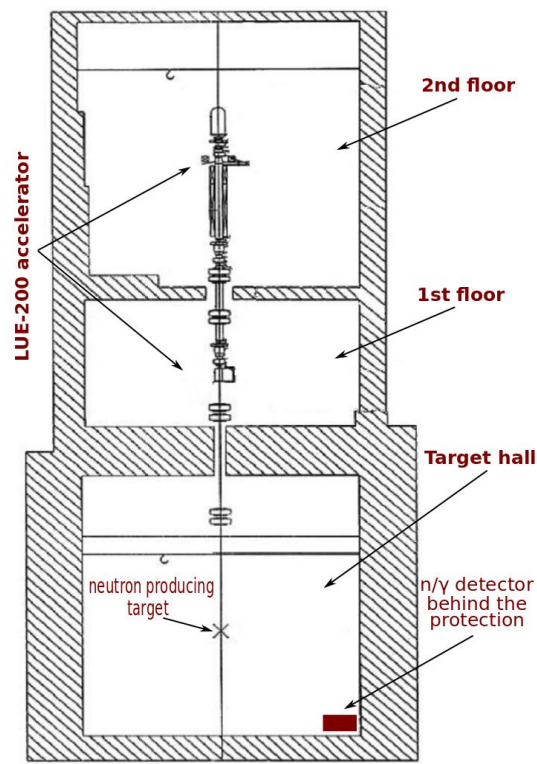

**Fig. 4.** Schema of the IREN facility and location of the neutron/gamma detector in the target hall.

A low-voltage power source of +6 V was used to bias the detector. The analog signal obtained from the detector was transferred to the preamplifier and subsequently fed to an analog-to-digital converter (ADC) with a resolution of 12 bits, specifically, the CAEN Desktop DT5720B. This digitizer was connected to a computer via USB. A specially written computer program in C++ allowed controlling and visualising the registration process in real time. The visualization program allowed monitoring the count rate changes over a long time. All data was saved to a computer for further analysis.

## 4. Experimental results
The power of the IREN facility is initiated by an electron pulse from the linear electron accelerator LUE-200. The accelerator has a pulse duration of 100 ns, an electron current per pulse of 1.5 A, and an electron energy of 30 MeV. The cycling of the accelerator is usually 25 Hz. Also, the accelerator can operate at different cycle frequencies (up to 50 Hz) [1-3].

The developed neutron/gamma detector is sensitive to changes in the operating mode of the facility. Fig. 5 shows the results of measuring the neutron count rate at different accelerator frequencies. The measurements were carried out with the accelerator turned off (background), at 2 Hz, 5 Hz, 10 Hz, 15 Hz, 20 Hz, 25 Hz, 30 Hz, 40 Hz, and 50 Hz, respectively. The duration of each measurement was 3 minutes. As can be seen from Fig. 5, the neutron/gamma detector count rate correlates very well with the operating mode of the IREN facility. Background events were 20 pulses per second at a threshold level of 10 mV.

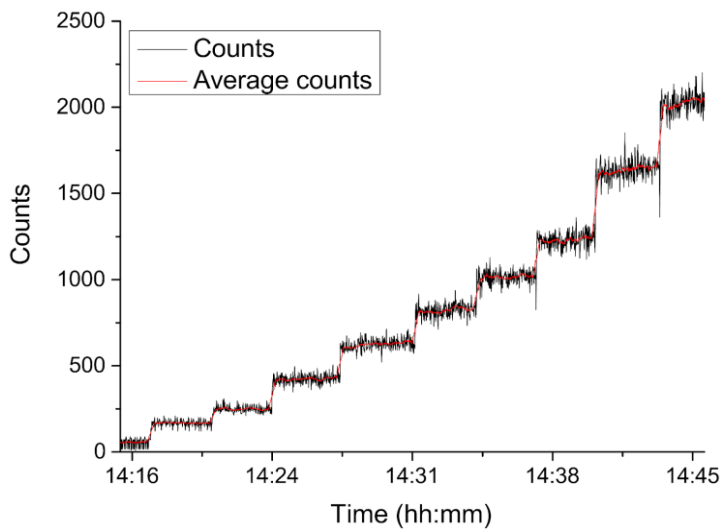

**Fig. 5.** Illustration of neutron/gamma count monitoring during the operation of the resonant neutron source IREN in the pulsed mode.

Fig. 5 demonstrates how the neutron detection count rate had a clear linear dependence on IREN neutron flux. An increase in the frequency of accelerator cycles correspondingly leads to an increase in the IREN neutron flux. It is expected that a linear increase in the IREN frequency of the accelerator correspondingly results in a proportional increase in the recorded counts detected by the detector. Here, a linear fit was applied to the detector raw count rate as a function of IREN neutron flux. Measurements were made at 10 different frequency levels of the accelerator, which correspond to different frequency values of the IREN facility (fig. 6). The dependence of the frequency on the counts, recorded by the detector with and without the background, showed that there were no additional (environmental) effects registered during the measurement.

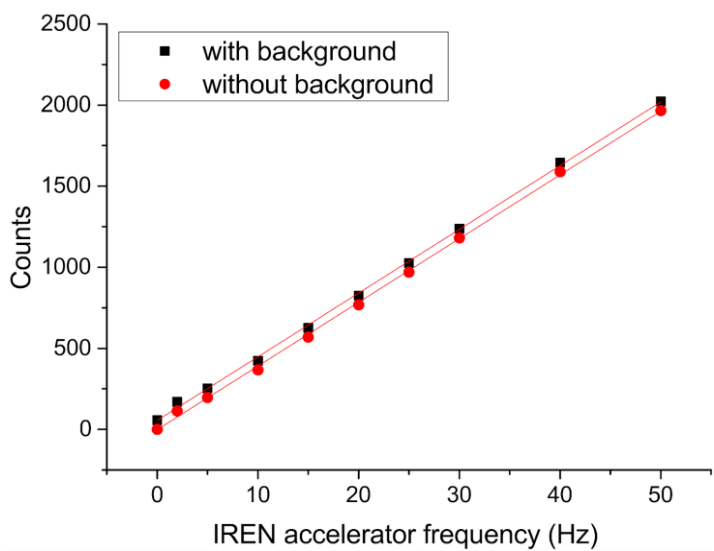

**Fig. 6.** Measured detector count rate versus IREN neutron flux, with a linear fit to demonstrate their linear dependence.

So it should be noted that detector count rate was not needed to be corrected for any environmental influences since changing atmospheric and/or environmental conditions leaded to small count rate fluctuations. The contribution of these factors was shown in fig. 7. As shown in fig. 7, the neutron/gamma counts practically did not change with time.

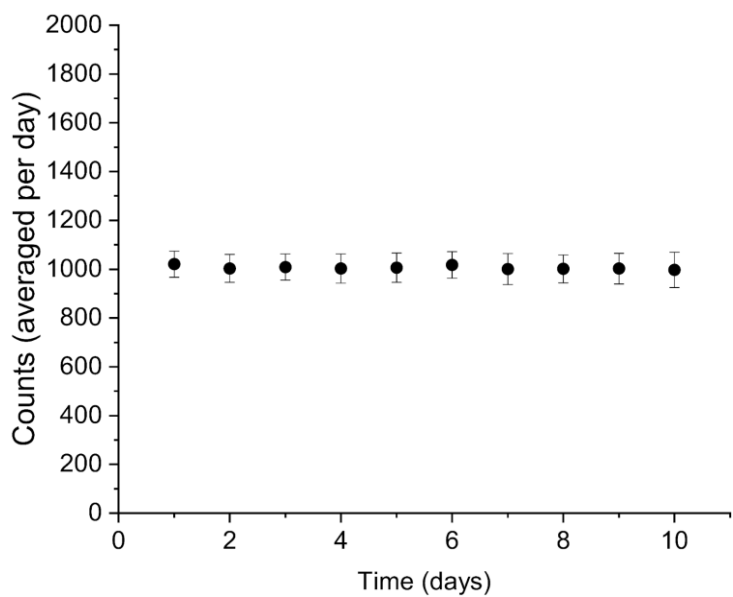

**Fig. 7.** Dependence of averaged counts per day on time. Measurements for 10 day were shown.

The highest rate of error recorded during measurement was 6 %. This was tested for 3 months. As known, breakdown voltage variation directly affects the operating voltage, which in turn causes a change in some of its technical parameters of MAPD [23]. Therefore, it is important to point out the main cause of the error was variation of temperature in target hall, which influences breakdown voltage of MAPD.

**5. Summary and conclusions**
The low-cost and portable neutron/gamma detector was designed and developed for status monitoring of IREN facility. The detector consists of MAPD with size of $3.7 \times 3.7$ mm$^2$ and the plastic scintillator with size of $3.7 \times 3.7 \times 30$ mm$^3$. The electronics for the detector was also designed and developed including the DC-DC converter (from 5 V up to 130 V), the preamplifier with 36 gain and the comparator with 10 mV minimum level. It was not necessary to measure the absolute neutron flux and discriminate signals from gamma and neutrons, since the detector module was developed for monitoring output neutron flux to control the status of the IREN facility. The threshold of the detector was according to the background level of the beam hall (the background gamma field is always present in the IREN target hall). The detector was sensitive to any power fluctuations. By monitoring the readings of the detector, an operator of the facility can make real-time adjustments to the mode of operation of the neutron source, while also maintaining control over the status of the IREN facility by comparing the results to those obtained from other monitors. The detector was tested at the IREN facility for 3 months and is running as a status monitoring device.

**Acknowledgments**
This project has received funding from the European Union's Horizon 2021 Research and Innovation Programme under the Marie Sklodowska-Curie grant agreement 101086178. We would like to thank Zecotek Photonics Inc. for providing MAPD samples.